\documentclass{procmult}
\usepackage{graphicx}

\begin{document}

\title{The Role of Time Gauge in Quantizing Gravity}
\author{Francesco Cianfrani$^\dag$, Giovanni Montani$^\ddag$}
\institute{
$^\dag$  ICRA-International Center for Relativistic Astrophysics, Physics Department (G9), University of Roma ``Sapienza'', Piazzale Aldo Moro 5, 00185 Rome, Italy. \\
$^\ddag$ ICRA-International Center for Relativistic Astrophysics, Physics Department (G9), University of Roma ``Sapienza'', Piazzale Aldo Moro 5, 00185 Rome, Italy.\\
ENEA C.R. Frascati (Dipartimento F.P.N.), via Enrico Fermi 45, 00044 Frascati, Rome, Italy.\\
ICRANET C. C. Pescara, Piazzale della Repubblica, 10, 65100 Pescara, Italy.
}
\maketitle
\abstract
We present the Hamiltonian formulation of General Relativity with the Holst formulation in a generic local Lorentz frame. In particular, we outline that a Gauss constraint is inferred by a proper generalization of Ashtekar-Barbero-Immirzi connections. This feature allow to extend the Loop Quantum Gravity quantization procedure to the case in which no gauge fixing at all is performed of the Lorentz frame.

\section{Introduction}

The definition of a rigorous quantum theory for the gravitational field is among the most important tasks of Theoretical physics. An intriguing point concerning a Quantum Gravity theory is the possibility that it would give some violations of fundamental symmetries, which opens important scenario for testing the proposed model (see for instance \cite{CSUV04}). Among fundamental symmetries, we focus our attention on the invariance under local boost transformations. 

We consider the Holst formulation \cite{Ho96} for gravity, for which the quantization procedure is well-defined at the kinematical level along the lines of Loop Quantum Gravity (LQG) \cite{revloop}. A proper feature of this approach consist in recognizing that, as soon as Ashtekar-Barbero-Immirzi connections \cite{ABI} are taken as configuration variables, the phase space resembles that of an SU(2) gauge theory. Hence, a proper Hilbert space can be defined by the quantization of the holonomy-flux algebra, using technique developed in the framework of lattice gauge theory \cite{lgt}. One of the most impressive results within this framework consists in the discreteness of geometrical operator spectra \cite{discr}, which outlines that the fundamental spatial structure is granular. 
In view of this discreteness, the investigation of the local Lorentz invariance is a very tantalizing subject. The preservation of this symmetry is non-trivial, since the whole LQG formulation is based on fixing before quantizing a particular local Lorentz frame adapted to the space-time slicing by the so-called time-gauge condition. 

The starting point for this analysis is the Hamiltonian formulation contained in the work by Barros e Sa \cite{BS01}, where the emergence of some second-class constraints and the possibility to solve them is outlined. This way, it is shown that the whole set of constraints can be reduced to a first-class one. This feature demonstrates that fixing the time-gauge is a well-grounded procedure. However, no indication is given whether the SU(2) Gauss constraints are preserved once a different choice is made for the local Lorentz frame. This is a crucial point in order to establish if the LQG quantization procedure works also without the time gauge.

Here, we provide a new solution for second-class constraints and we demonstrate \cite{prl} that in this scheme there is no need of fixing the local Lorentz frame in view of finding the phase space structure proper of LQG, {\it i.e.} that of an SU(2) gauge theory. Furthermore, the invariance under boost can be imposed by an additional operator, which in a suitable set of phase-space coordinates give simply the condition that physical states do not depend on the variables $\chi_a$ labeling different frames. Such constraints can be solved after quantizing the dynamical systems, thus showing that the boost invariance can be implemented on a quantum level.   

As a consequence, no modification arises with respect to the case when the time gauge holds. For instance we investigate the area operator and we outline that the corresponding spectrum does not depend on the local Lorentz frame.

The organization of the manuscript is as follows: in section \ref{1} the Hamiltonian formulation of the Holst action is reviewed and the solution to second-class constraints is provided in section \ref{2}. Then, in section \ref{3} it is demonstrated that SU(2) Gauss constraints appear in a time-independent frame, while in section \ref{6} the extension to time-dependent frames is addressed. Hence, the LQG formulation is applied in a framework without any gauge fixing in section \ref{4}. Finally, brief concluding remarks follow in \ref{5}.         

\section{Hamiltonian formulation}\label{1}
We start with the Holst action \cite{Ho96} for gravity, which in units $8\pi G=c=1$ reads as follows
\begin{equation} 
S=\int \sqrt{-g}e^\mu_A e^\nu_B R_{\mu\nu}^{CD}(\omega^{FG}_\mu){}^\gamma\!p^{AB}_{\phantom1\phantom2CD},\label{act}
\end{equation}

$g$ being the determinant of the metric tensor $g_{\mu\nu}$ with 4-bein vectors $e^A_\mu$ and spinor connections $\omega^{AB}_\mu$, while the expressions for $R^{AB}_{\mu\nu}$ and ${}^\gamma\!p^{AB}_{\phantom1\phantom2CD}$ are
\begin{equation}
R^{AB}_{\mu\nu}=\partial_{[\mu}\omega^{AB}_{\nu]}-\omega^A_{\phantom1C[\mu}\omega^{CB}_{\nu]},\qquad{}^\gamma\!p^{AB}_{\phantom1\phantom2CD}=\delta^{AB}_{\phantom1\phantom2CD}-\frac{1}{2\gamma}\epsilon^{AB}_{\phantom1\phantom2CD}. \label{EH}
\end{equation}

$\gamma$ is the Immirzi parameter, the fundamental ambiguity arising in Loop Quantum Gravity. Conjugate momenta ${}^\gamma\!\pi_{AB}^\mu$ are given by
\begin{equation} 
{}^\gamma\!\pi^t_{AB}=0,\qquad {}^\gamma\!\pi_{AB}^i={}^\gamma\!p^{CD}_{\phantom1\phantom2AB}\pi_{CD}^i,
\end{equation}

where $\pi^i_{AB}=2\sqrt{-g}e^t_{[A}e^i_{B]}$. 

At this level the following constraints are obtained 
\begin{equation}\pi_{AB}^t=0,\qquad C^{ij}=\epsilon^{ABCD}\pi_{AB}^{(i}\pi_{CD}^{j)}=0,\end{equation}

and the expression of the Hamiltonian is
\begin{eqnarray}   \mathcal{H}=\int\bigg[\frac{1}{eg^{tt}}H-\frac{g^{ti}}{g^{tt}}H_i-\omega^{AB}_t{}^\gamma\!p^{CD}_{\phantom1\phantom2AB}G_{CD}+\lambda_{ij}C^{ij}+\eta_{ij}D^{ij}+\lambda^{AB}\pi_{AB}^t\bigg]d^3x,\end{eqnarray}

where $1/eg^{tt}$, $g^{ti}/g^{tt}$, ${}^\gamma\!p^{CD}_{\phantom1\phantom2AB}\omega^{AB}_t$, $\lambda_{ij}$, $\eta_{ij}$ and $\lambda^{AB}$ behave as Lagrangian multipliers. Hence, on the subspace $\{\omega^{AB}_i,\pi^i_{AB}\}$ the constraint hypersurfaces is given by
\begin{equation}   
\left\{\begin{array}{c}H=\pi^i_{CF}\pi^{jF}_{\phantom1D}{}^\gamma\!p^{CD}_{\phantom1\phantom2AB}R^{AB}_{ij}=0 \\\\ 
H_i={}^\gamma\!p_{AB}^{\phantom1\phantom2CD}\pi^j_{CD}R^{AB}_{ij}=0 \\\\ G_{AB}=D_i\pi^i_{AB}=\partial_i\pi^i_{AB}-2\omega_{[A\phantom2i}^{\phantom1C}\pi^i_{|C|B]}=0 \\\\ C^{ij}=\epsilon^{ABCD}\pi_{AB}^{(i}\pi_{CD}^{j)}=0 \\\\ D^{ij}=\epsilon^{ABCD}\pi^k_{AF}\pi^{(iF}_{\phantom1\phantom2B}D_k\pi^{j)}_{CD}=0
\end{array}\right.,
\end{equation}

where $D^{ij}=0$ arise as secondary constraints from $C^{ij}=0$. The full set of constraints is second-class, because the Poisson brackets $\{C^{ij},D^{kl}\}$ and $\{D^{ij},D^{kl}\}$ do not vanish on the constraints hypersurfaces.

The physical interpretation of such constraints is clear: $H$ and $H_i$ are the super-Hamiltonian and the super-momentum, respectively, and they ensure the invariance under diffeomorphisms in the ADM representation,  $G_{AB}$ are Gauss constraints of the Lorentz symmetry. As far as $C^{ij}$ and $D^{ij}$ are concerned, they made the whole system second-class, thus they are not associated with any gauge symmetry.

\section{Solution of second-class constraints}\label{2}
The solution to $C^{ij}=D^{ij}=0$ is given by
\begin{eqnarray}
\pi^i_{ab}=2\chi_{[a}\pi^i_{b]},\label{scon1}\\ \omega^{\phantom1b}_{a\phantom1i}={}^\pi\!\omega^{\phantom1c}_{a\phantom1i}T^{-1b}_c(\chi)+\chi_a\omega^{0b}_{\phantom{12}i}+\chi^b
(\omega_{a\phantom1i}^{\phantom10}-\partial_i\chi_a),\label{scon2}
\end{eqnarray}

$\chi_a$ being arbitrary functions of space-time coordinates.  ${}^\pi\!\omega^{\phantom1b}_{a\phantom1i}$ can be thought as connections associated to $\pi^i_a$, in fact their expression reads 
\begin{equation}
{}^\pi\!\omega^{\phantom1b}_{a\phantom1i}=\frac{1}{\pi^{1/2}}\pi^b_l{}^3\!\nabla_i(\pi^{1/2}\pi^l_a), 
\end{equation}

$\pi^a_i$ being the inverses of $\pi^i_a$, while ${}^3\!\nabla_i$ is the covariant derivative associated to the metric $h_{ij}=-\frac{1}{\pi}T^{-1}_{ab}\pi^a_i\pi^b_j$. We denote by $\Omega$ the hypersurfaces where conditions (\ref{scon1}) and (\ref{scon2}) hold.

The geometrical interpretation of functions $\chi_a$ can be given starting from a $3+1$ splitting of the space-time manifold. In fact, as soon as time-like and space-like coordinates are identified (we denote them by $t$ and $x^i$, $i=1,2,3$, respectively), 4 bein vectors can be decomposed as follows
\begin{equation}
e^0_\mu=(N, \chi_a E^a_i),\qquad e^a_\mu=(E^a_iN^i, E^a_idx^i).\label{tetr}
\end{equation} 

It can be shown from the definition of $\pi^i_{AB}$ that $\chi_a$ into the solutions (\ref{scon1}) and (\ref{scon2}) coincide with the ones introduced in the expression above.

Therefore, the vector $\chi_a$ gives the velocity of the frame $\{e^a\}$ with respect to spatial hypersurfaces \cite{FG07}. 

Furthermore, within this scheme the following correspondence between ADM variables and components of the 4-bein holds
\begin{equation}
\widetilde{N}=\frac{N-N^i\chi_aE^a_i}{1+\chi^2},\qquad \widetilde{N}^i=N^i-\frac{N-N^i\chi_bE^b_i}{1+\chi^2}\chi^aE_a^i,
\end{equation}

$\widetilde{N}$ and $\widetilde{N}^i$ being the lapse function and the shift vector, while $\chi^a=\eta^{ab}\chi_b$ and $\chi^2=\chi^a\chi_a$. 
Moreover $E^a_i=\frac{1}{1+\chi^2}\pi_i^a$ and the metric of spatial hypersurfaces is $h_{ij}$ itself. 

For later purposes we introduce inverse densitized 3-bein vectors $\widetilde{\pi}^i_a$, which can be expressed in terms of $\pi^i_a$ as follows
\begin{equation}
\widetilde{\pi}^i_a=S^b_a\pi^i_b,\qquad S_a^b=\sqrt{1+\chi^2}\delta^a_b+\frac{1-\sqrt{1+\chi^2}}{\chi^2}\chi_a\chi_b.
\end{equation} 
 
\section{Time-independent local Lorentz frame}\label{3}
Let us now focus our attention to the case $\partial_t\chi_a=0$, {\it i.e.} we treat $\chi_a$ as fixed parameters.
We can take as coordinates on $\Omega$ the variables $\{\omega^{0a}_i,\pi^i_a=\pi^i_{0a}\}$. However, since  second-class constraints have been solved, the induced symplectic form is non-trivial. In particular, it is given by
\begin{eqnarray}
\{\pi_a^i(x,t),\pi_b^j(y,t)\}=0\qquad\qquad\qquad\qquad\qquad\qquad\qquad\qquad\qquad\qquad\qquad\label{1c}\\
\{\omega^{a0}_i(x,t),\omega^{b0}_j(y,t)\}=\left(-\frac{1}{2\gamma(1+\chi^2)^2}T^{-1a}_{\phantom1c}T^{-1b}_{\phantom2d}T^{-1g}_{\phantom1h}\epsilon^d_{\phantom1fg}-\frac{1}{1+\chi^2}T^{-1a}_{\phantom1c}\chi_h\delta_f^b\right)\frac{\partial{}^\pi\!\omega^{fh}_j(y,t)}{\partial\pi^i_c(x,t)}-\nonumber\\-\left(-\frac{1}{2\gamma(1+\chi^2)^2}T^{-1b}_{\phantom1c}T^{-1a}_{\phantom2d}T^{-1g}_{\phantom1h}\epsilon^d_{\phantom1fg}-\frac{1}{1+\chi^2}T^{-1b}_{\phantom1c}\chi_h\delta_f^a\right)\frac{\partial{}^\pi\!\omega^{fh}_i(x,t)}{\partial\pi^j_c(y,t)}\qquad\label{2c}\\
\{\omega^{a0}_i(x,t),\pi_b^j(y,t)\}=\delta^j_i\delta^3(x-y)\frac{1}{1+\chi^2}T^{-1a}_{\phantom1b}.\qquad\qquad\qquad\qquad\qquad\qquad\qquad
\label{3c}\end{eqnarray}

In view of relations (\ref{scon1}), (\ref{scon2}) $G_{AB}$ can be rewritten as 
\begin{equation}
\left\{\begin{array}{c} G_{ab}=2(1+\chi^2)T^c_{[a}T^d_{b]}(\omega_{0ci}+{}^\pi\!D_i\chi_c)\pi^i_d \\ 
G_{0a}=2\chi^b(\omega_{0[ai}+{}^\pi\!D_i\chi_{[a})\pi^i_{b]} \end{array}\right.,
\end{equation}

where ${}^\pi\!D_i\chi_{a}=\partial_i\chi_a-{}^\pi\!\omega^{\phantom1b}_{a\phantom1i}\chi_b$.

The relations above emphasize that $G_{AB}=0$ are not independent conditions, but we have $G_{0a}=\chi^bG_{ab}$.

Therefore, 3 constraints become redundant and we perform our analysis on the remaining conditions $G_{ab}=0$.
However, the algebra of constraints $G_{ab}=0$ turns out to be open. Nevertheless, it is possible to sum up the following vanishing contribution
\begin{equation}
\partial_i\pi^i_a-{}^\pi\!\omega_{a\phantom1i}^{b}\pi^i_b-\frac{\partial_i\chi^2}{2(1+\chi^2)}\pi^i_a,
\end{equation}

so finding 
\begin{equation}
\partial_i\pi^i_a+\frac{1}{2}\frac{\partial_i\chi^2}{1+\chi^2}\pi^i_a-\frac{1}{2}\partial_i(\chi_a\chi^b)T^{c}_b\pi^i_c-\gamma\epsilon_{ab}^{\phantom{12}c}A^b_iT_c^d\pi_d^i=0,\label{bal}
\end{equation}

where $A^a_i$ is given by
\begin{equation}
A_i^a=(1+\chi^2)T^{ac}(\omega_{0ci}+{}^\pi\!D_i\chi_c)-\frac{1}{2\gamma}\epsilon^a_{\phantom1cd}{}^\pi\!\omega^{cf}_{\phantom1\phantom2i}T^{-1d}_{\phantom1f}.\label{acon}
\end{equation}

If we multiply the expression (\ref{bal}) times $S^a_b$, we find  
\begin{eqnarray}
G_a=\partial_i\widetilde{\pi}^i_a+\gamma\epsilon_{ab}^{\phantom{12}c}\widetilde{A}^b_i\widetilde{\pi}_c^i=0,\label{rcon}
\end{eqnarray}

$\widetilde{A}^a_i$ being
\begin{equation}
\widetilde{A}_i^a=S^{-1a}_b\left(A^b_i+\frac{2+\chi^2-2\sqrt{1+\chi^2}}{2\gamma\chi^2}\epsilon^{abc}\partial_i\chi_b\chi_c\right).\label{ABI}
\end{equation}

Therefore, \emph{SU(2) Gauss constraints are inferred also without the time-gauge condition}.

Furthermore, SU(2) connections $\widetilde{A}^a_i$ and densitized inverse 3-bein vectors $\widetilde{\pi}^i_a$ are a couple of canonically conjugate variables, since it can be shown from relations (\ref{1c}), (\ref{2c}) and (\ref{3c}) that
 $\{\widetilde{A}^a_i(t,x),\widetilde{\pi}^j_b(t,y)\}=\delta^a_b\delta^{j}_i\delta^3(x-y)$, while other Poisson brackets vanish. 

These features allow us to conclude that $\widetilde{A}^a_i$ are \emph{the extension of Barbero-Immirzi connections to a generic time-independent Lorentz frame} and that the phase space of General Relativity resembles that of an SU(2) gauge theory also when the time-gauge condition does not hold.
As far as other constraints are concerned, {\it i.e.} the super-momentum and the super-Hamiltonian, since they are invariant under transformations of the local Lorentz frame, their expression in terms of $\widetilde{A}^a_i$ and $\widetilde{\pi}^i_a$ is the same as in LQG with the time gauge.  

\section{On the generalization to time-dependent frames}\label{6}

The extension of the proposed procedure to the case $\partial_t\chi_a\neq0$ requires to take into account the dynamical role played by $\chi_a$ themselves, {\it i.e.} conjugate momenta $\pi^a$ must be added. We introduce $\pi^a$ by imposing that the initial set of phase space variables $\{\omega_i^{AB},{}^\gamma\!\pi^j_{CD}\}$ is mapped into $\{\widetilde{A}^a_i,\chi_b,\widetilde{\pi}_c^j,\pi^d\}$ by a canonical transformation, {\it i.e.}. 
\begin{equation}
\frac{1}{2}{}^\gamma\!\pi^i_{AB}\partial_t\omega^{AB}_i=\widetilde{\pi}^i_a\partial_t\widetilde{A}^a_i+\pi^a\partial_t\chi_a.
\end{equation}

Such a request leads to the following conditions
\begin{equation}
T^{-1a}_b\pi^b=0.\label{picon}
\end{equation}

The emergence of such additional constraints is not surprising, since we enlarged the set of phase space coordinates by treating $\chi_a$ as configuration variables, hence the redundancy of some variables is expected. In particular, we discussed in section \ref{3} that as soon as the solution of second-class constraints are inserted into the Gauss constraints of the Lorentz group, only three independent conditions remain. Constraints (\ref{picon}) are the conditions one must add to maintain the same number of degrees of freedom, thus they replace the Gauss constraints associated with the boost part of the Lorentz group.  

Therefore, \emph{conjugate momenta to $\chi_a$ are constrained to vanish}. 

This result can be regarded as outstanding, since it demonstrates that the Gauss constraints of the Lorentz group, which involve a mixing of two set of conjugate variables (connections and $\chi_a$), is equivalent to two set of constraints (\ref{rcon}) and (\ref{picon}), each one acting on a single set of conjugate variables.

Finally, the full action can be written as
\begin{equation}
S=\int d^4x\bigg[\widetilde{\pi}^i_a\partial_t\widetilde{A}^a_i+\pi^a\partial_t\chi_a-\frac{1}{\sqrt{g}g^{tt}}H+\frac{g^{ti}}{g^{tt}}H_i+({}^\gamma\!p^{cd}_{\phantom{12}AB}+2\chi^d{}^\gamma\!p^{0c}_{\phantom{12}AB})\omega^{AB}_t\epsilon^b_{\phantom1cd}S^{-1a}_bG_a+\lambda^aB_a\bigg],\label{FINALACT}
\end{equation}

$\lambda^a$ being Lagrangian multipliers. In this context rotations and boosts are generated by
\begin{equation}
\left\{\begin{array}{c} R_a=G_a+\epsilon_a^{\phantom1bc}\chi_bB_c, \\  K_a=B_a+\frac{1-\sqrt{1+\chi^2}}{\chi^2}\epsilon_a^{\phantom1bc}G_b\chi_c \end{array}\right.,\label{Lt}
\end{equation}

respectively. These relations can be inverted and this implies that the validity of the Hamiltonian constraints is equivalent to the invariance under the action of the Lorentz group.

\section{LQG in a generic local Lorentz frame.}\label{4}

The results of the previous sections turn out to be very useful in view of extending the quantization procedure of LQG to the case in which no gauge fixing of the Lorentz frame is performed. The main difficulty one encounters when the full Lorentz symmetry is explicitly preserved consists in the fact that the associated group of transformations is non-compact. As a consequence we cannot use standard techniques for the quantization of gauge theories, since some divergences arise when performing the integration on the group manifold via the Haar measure. This is the main reason for the failure to quantize directly a gauge theory of the Lorentz group. 

Here, the Gauss constraints of the Lorentz group are replaced by two set of constraints, which can be safely imposed following the Dirac procedure for the quantization of constrained systems. In fact, the condition that $B^a$ annihilates physical states, {\it i.e.} 
\begin{equation}
\hat{B}_a\psi(\chi_b,\widetilde{A}^c_i)=0,
\end{equation}

can be solved in the most natural operator ordering taking wave functionals not depending on $\chi_a$ variables. 

As for $\hat{G}_a$, they are Gauss constraints of the SU(2) group and the standard LQG quantization procedure works without modifications. 

Let us define holonomies $h(\widetilde{A})_\alpha$ along an edge $\alpha=\alpha(t)$ and fluxes $\widetilde{\pi}(S)$ across a surface $S=S(u,v)$ as follows  
\begin{eqnarray}
h(\widetilde{A})_\alpha=Pe^{\int_0^1\widetilde{A}^a_i\dot{\alpha}dt\tau_a}\\ 
\widetilde{\pi}(S)_a=\int \widetilde{\pi}^i_an_idudv,
\end{eqnarray}

$n_i$ being the normal vector to $S$, while $\tau_a$ are SU(2) generators.

The holonomy-flux algebra is the same one as in LQG with the time gauge, so the quantization can be performed in the same way. 

The representation on a Hilbert space of cylindrical functions $f_\alpha(g_1,\ldots,g_N)$ is obtained by the GNS construction, taking the following ground state $\omega$ \cite{LOST06}
\begin{equation}
\omega(f_\alpha(\widetilde{A}),\widetilde{\pi}_1,..,\widetilde{\pi}_M)=\left\{\begin{array}{cc}\int d\mu(g_1)..d\mu(g_N)f_\alpha(g_1,..,g_N) & M=0 \\ 0 & M\neq0 \end{array}\right.,
\end{equation}

where the integration is performed via the Haar measure for the SU(2) group. 

This way the Hilbert space is a certain completition over the space of distributional connections
\begin{equation}
\textsc{H}_{aux}=L^2(\it{\bar{A}},d\mu),
\end{equation}

whose scalar product is defined by the Ashtekar-Lewandowsky measure \cite{ALMMT95}. A basis for this space is given by invariant spin networks, which on a piece-wise analytic graph $\alpha$ can be written as
\begin{equation}
\psi_{\alpha}(\widetilde{A})=\left(\prod_{v_\alpha\in k}I_{v}\right)\left(\prod_{e_\alpha\in k}D_{\rho_{e}}\rho_{e}(h_e(\widetilde{A}))\right),
\end{equation}

$v_\alpha$ and $e_\alpha$  being verticies and edges of $\alpha$, respectively, while $I_v$ are invariant inter-twiners and $\rho_{e}(h_e(\widetilde{A}))$ is the $D_{\rho_{e}}$-dimensional irreducible SU(2) representation associated with the holonomy along the edge $e$. 

In view of relations (\ref{Lt}) physical states developed as SU(2) gauge-invariant ones and not depending on $\chi_a$ must be scalar under Lorentz transformations. Therefore, the full Lorentz symmetry is preserved on the physical sector of the quantum theory.

Within this scheme, since the geometrical meaning of variables is the same as when the time gauge holds, no modification is expected with respect to the standard LQG formulation.  

For instance, let us consider the area operator $A$, whose eigen-functionals are invariant spin-networks when the time gauge holds. Given a surface $S$ and an edge $e$ having one intersection in a point $p$ (where the tangent to $e$ does not belong to $S$), the action of the operator $A(S)$ on the parallel transport along $e$ is given by
\begin{equation}
A(S)h_e(A)=\gamma\sqrt{1+\chi^2}\sqrt{O}h_e(A),\qquad O=\frac{1}{1+\chi^2}\eta^{ab}\tau_a\tau_b,\label{area}
\end{equation}

$\tau_a$ being $SU(2)$ generators. From the relation above it can be shown that invariant spin-networks are still eigen-functionals of the area operator and that the spectrum is the same as in standard LQG.  

Thus \emph{the area spectrum is discrete and this discrete structure does not depend on $\chi_a$}. 

This result has a relevant physical meaning, but it can be regarded as no surprising as far as we make a comparison with the rotation invariance and the discrete spectrum of the angular momentum operator in ordinary quantum mechanics.

\section{Conclusions}\label{5}
We reviewed the Hamiltonian formulation of the Holst action and we provided a new solution of second-class constraints. Then, we analyzed the phase space structure on the hypersufaces where these constraints hold in the case of a time-independent Lorentz frame. We found that SU(2) Gauss constraints can be recovered. This way, we generalized Ashtekar-Barbero-Immirzi connections. Furthermore, we recognized that the same Gauss constraints appear also when no restriction to time-independent frames was performed. However, in this case additional constraints came out from the requirement of dealing with a canonical map from Lorentz connections to SU(2) ones. Nevertheless, such constraints implied simply the non-dynamical character of the variables $\chi_a$, giving the velocity components of the Lorentz frame with respect to spatial hypersurfaces. In this respect they can be easily solved taking wave-functional not depending on $\chi_a$. This shows that the boost invariance can be realized on a quantum level. 

As a confirmation of this scenario, we carried on the quantization of gravity following LQG and we demonstrated that the spectrum of the area operator did not change.

Therefore, fixing the time-gauge condition is not a necessary tool to quantize gravity, since by using a proper set of phase space variables the SU(2) gauge structure can be find out, while additional constraints can be solved on a quantum level, too.

This result shows that the granular spatial structure is invariant under tangent space transformations and it seem to suggest that no violation of the Lorentz symmetry is predicted in LQG. However, the absence of a proper semi-classical limit do not allow us to infer any conclusion regarding low-energy phenomena.  

The introduction of matter fields is an interesting extension of this framework. In particular, Dirac spinors deserve further investigations for two reasons
\begin{itemize}
{\item they modify second-class constraints, thus a new solution should be found,}
{\item they are coupled to Lorentz connections and we expect they to enter as sources into the SU(2) Gauss constraints.}  
\end{itemize}

\end{document}